\documentstyle[12pt,aps,epsfig,prl]{revtex}

\relax  
\begin{document}
\preprint{prl}
\draft

\title{Interlayer electrodynamics in the organic superconductor $\kappa-$(BEDT$-$TTF)$_2$Cu(NCS)$_2$: evidence for a transformation within the vortex state}

\bigskip

\author{S. Hill\cite{email}}
\address{Department of Physics, University of Florida, Gainesville, FL 32611}

\author{M. M. Mola\cite{address}}
\address{Department of Physics, Montana State University, Bozeman, MT 59717}

\author{J. S. Qualls}
\address{Department of Physics, Wake Forest University, Winston-Salem, NC 27109}

\date{\today}
\maketitle

\bigskip

\begin{abstract}

A microwave cavity perturbation technique is used to probe the
interlayer electrodynamics within the vortex state of the organic
superconductor $\kappa-$(BEDT$-$TTF)$_2$Cu(NCS)$_2$. A Josephson
plasma mode is observed which is extremely sensitive to
correlations in the locations of vortices in adjacent layers and
may, therefore, be used to gauge collective effects between
vortices and crystal pinning sites in the title compound. Our
previous investigations [M. M. Mola {\em et al.}, Phys. Rev. B
{\bf 62} (2000) 5965] revealed a transformation from a correlated
quasi-two-dimensional pinned vortex phase, to either a depinned or
liquid state. In this study, we carry out a detailed analysis of
the magnetic field dependence of the Josephson plasma frequency
within the two phases. Our findings agree favorably with recent
theoretical models: within the liquid state, the squared plasma
frequency ($\omega_p^2$) decays with the inverse of the magnetic
field strength, {\bf B}; whereas, in the pinned phase, a much
slower decay is observed ($\omega_p^2\propto$ {\bf B}$^{-0.35}$),
which is indicative of weak pinning.

\end{abstract}

\bigskip

%\bigskip

%\begin{multicols}{2}[]
\clearpage \centerline{\bf I. Introduction}
\bigskip

\noindent{The equilibrium magnetic field$-$temperature phase
diagram for type-II superconductors is extremely rich, including
many different vortex solid, liquid and glassy phases
\cite{Crabtree,BlatterRMP}. Presently, the theory of vortex
interactions in layered superconductors is incomplete, though much
work has focused on this problem in recent years
\cite{Crabtree,BlatterRMP,Bul95,Bul96,Koshelev96,Koshelev98,Tsui,Matsuda,Zeldov,Schilling}.
Information concerning vortex structure and dynamics is important
for two reasons: first, the vortex structure contains important
information regarding the symmetry of the superconducting state;
second, vortex motion leads to dissipation, and an understanding
of the dissipative mechanisms in superconductors is essential for
progress in developing viable technology based on these materials.
Superconducting vortices also provide an excellent laboratory for
general phase-transformation behavior \cite{Crabtree}.
Experimentally, all of the relevant parameters can be varied over
wide ranges: the vortex density by many orders of magnitude by
changing the magnetic field; thermal fluctuations by varying the
temperature; pinning by varying disorder; and, in quasi-two
dimensional (Q2D) systems, the coupling between the layers may be
varied through the choice of material.}

In the layered cuprate high temperature superconductors (HTS) and,
more recently, in the Q2D organic superconductors (OS), it is
widely accepted that Josephson coupling is responsible for the
interlayer transport of Cooper pairs \cite{Kleiner}. This adds to
the complexity of the mixed state phase diagram $-$ pancake
vortices threaded by the same magnetic flux quantum may become
completely decoupled from one layer to the next, creating a Q2D
vortex state \cite{Crabtree,BlatterRMP}. A model system for
investigations of Q2D vortex physics, and one which has received
relatively little attention in comparison with the HTS, is the 10
K organic superconductor $\kappa-$(BEDT-TTF)$_2$Cu(NCS)$_2$, where
BEDT-TTF denotes bis-ethylenedithio-tetrathiafulvalene
\cite{Ishiguro,Single1}. Like the oxide HTS,
$\kappa-$(BEDT-TTF)$_2$Cu(NCS)$_2$ possesses a layered structure
in which highly conducting BEDT-TTF planes are separated by
insulating anion layers; for this material, the least conducting
direction is along the crystallographic $a-$axis \cite{Ishiguro}.
The anisotropy parameter in the normal state, given by the ratio
of the in-plane to out-of-plane conductivities
$\sigma_{bc}/\sigma_a$, is $\sim 1000$. In the superconducting
state, the anisotropy parameter is given by $\gamma \equiv
\lambda_\perp/\lambda_\parallel \sim 100-200$, where $\lambda
_\parallel$ and $\lambda _\perp$ are the London penetration depths
for AC currents induced parallel ($\lambda _\parallel \sim 0.8 \mu
m$) and perpendicular ($\lambda _\perp \sim 100 \mu m$) to the
conducting layers respectively
\cite{Ishiguro,Single1,JPR1,MolaPRL}. Such a large anisotropy
makes this OS a prime candidate to study Josephson coupling, and
changes in this coupling upon the introduction of vortices into
the sample, through the application of an external magnetic field.
Unlike many of the HTS, this material is extremely clean,
possessing far fewer crystal defects or pinning sites for magnetic
flux. Furthermore, because of the reduced T$_c$ and H$_{c2}$ (T$_c
= 10$ K and $\mu_o$H$_{c2} = 4$ tesla for the field perpendicular
to the layers), one can probe much more of the magnetic
field$-$temperature parameter space within the superconducting
state than is currently possible for the HTS
\cite{Ishiguro,Single1}.

Below T$_c$, a Josephson plasma resonance (JPR) dominates the
interlayer ($a-$axis) electrodynamics of
$\kappa-$(BEDT-TTF)$_2$Cu(NCS)$_2$
\cite{Shib1,Shib2,JPR1,Comment}. The squared JPR frequency
($\omega_p^2$) is directly proportional to the maximum interlayer
(or Josephson) current density $J_m(${\bf B},T). In turn,
$J_m(${\bf B},T) is related to the zero field interlayer critical
current density $J_o$ through the expression

\bigskip

\centerline{\hfill\hfill\hfill\hfill\hfill\hfill\hfill $J_m(${\bf
B},T) = $J_o\langle\langle \cos \varphi_{n,n+1}(${\bf
r}$)\rangle_t\rangle_d ,\hfill\hfill\hfill\hfill\hfill\hfill (1)
\hfill$}

\bigskip

\noindent{where $\varphi_{n,n+1}(${\bf r}) is the gauge-invariant
difference in the phase of the superconducting order parameter
between layers {\em n} and $n + 1$ at a point {\bf r} $ = x,y$ in
the $bc-$plane, and $\langle\cdot\cdot\cdot\rangle_t$ and
$\langle\cdot\cdot\cdot\rangle_d$ denote thermal and disorder
averages \cite{Bul96}. $\varphi_{n,n+1}(${\bf r}) depends
explicitly on the vortex structure within the mixed state and is,
thus, responsible for the field dependence of the JPR frequency
$\omega_p$. In the case of a 3D ordered flux-line-lattice, at T$ =
0$, $\langle\langle\cos\varphi_{n,n+1}(${\bf
r}$)\rangle_t\rangle_d = 1$ and maximum Josephson coupling occurs.
However, in the presence of any disorder, the flux lines will
deviate from linearity and
$\langle\langle\cos\varphi_{n,n+1}(${\bf r}$)\rangle_t\rangle_d$
will be suppressed. Sources of disorder include: crystal defects,
which create random vortex pinning sites; thermal fluctuations
which may lead to a vortex lattice melting transition; or a 3D to
2D crossover transition, whereby the flux lines lose their
rigidity, and the pancake vortices in adjacent layers decouple. A
comprehensive account of the influence of vortices and the role of
disorder is beyond the scope of this article; the interested
reader should refer to {\em e.g.} refs.
\cite{Bul95,Bul96,Koshelev96,Koshelev98}.}

In a previous investigation \cite{JPR1}, we reported an unusual
magnetic field dependence of the JPR frequency $\omega_p$, which
had been predicted for a weakly pinned Q2D pancake vortex phase.
In addition, we identified a possible transformation from this
pinned phase, to a liquid state, upon crossing the irreversibility
line (high-{\bf B}, high-T side) \cite{JPR1,MolaPRL}. In this
article, we extend the scope of these investigations. In
particular: we carefully evaluate the field dependence of
$\omega_p$ in the low {\bf B}/T pinned phase, and compare our
findings with detailed theoretical predictions; we carefully map
out the phase boundary separating the pinned and liquid phases
from measurements covering an extended frequency range ($16-200$
GHz); and we evaluate the field dependence of $\omega_p$ in the
high temperature phase, and consider various models for this
phase.

\bigskip

\centerline{\bf II. Experimental details}

\bigskip

\noindent{The high degree of sensitivity required for single
crystal measurements is achieved using a resonant cavity
perturbation technique in combination with a broad-band
Millimeter-wave Vector Network Analyzer (MVNA) exhibiting an
exceptionally good signal-to-noise ratio \cite{mmrsi}. The MVNA is
a phase sensitive, fully sweepable (8 to 350 GHz), superheterodyne
source/detection system. Several sample probes couple the network
analyzer to a range of high sensitivity cavities ($Q-$factors of
up to 25,000) situated within the bore of a 7 tesla
superconducting magnet. Current capabilities allow single crystal
measurements at any frequency in the range from 8 to 200 GHz, at
temperatures down to 1.5 K ($\pm0.01$K), and for any geometrical
combination of DC and AC field orientations up to 7 T (up to 45 T
at the National High Magnetic Field Laboratory); this
instrumentation is described in detail in ref. \cite{mmrsi}. The
use of a narrow band cavity offers many important advantages over
non-resonant methods. Careful consideration concerning the
coupling of radiation to and from the cavity (via waveguide),
combined with the ability to study very small samples, eliminates
problems associated with standing waves in the sample probe
\cite{mmrsi}. This, in turn, eliminates a mixing of the
dissipative and reactive responses of the sample under
investigation and, when combined with a vector detection scheme,
enables faithful extraction of both components of the complex
conductivity. Finally, the use of a cavity enables positioning of
a single crystal sample into a well defined electromagnetic field
environment, {\em i.e.} the orientations of the DC and AC magnetic
fields relative to the sample's crystallographic axes are
precisely known. In this way, one can systematically probe each
diagonal component of the conductivity tensor (in principle, the
off diagonal components also) \cite{lambda}.}

Within the superconducting state, dissipation is governed by the
surface resistance of the sample, {\em i.e.} the real part of the
surface impedance $\hat{Z} = R_S + i X_S = (i \mu_o
\omega/\hat{\sigma})^{1/2}$ \cite{lambda}. In this article,
measurements are restricted to geometries which probe only the
interlayer electrodynamics; the precise details as to how we
achieve this are described elsewhere \cite{mmrsi,lambda}.
Consequently, the measured dissipation depends only on the
interlayer conductivity $\hat{\sigma}_a(\omega$,{\bf B},T), which
includes contributions from the Josephson tunneling of Cooper
pairs and the normal quasiparticles. A simple two fluid model
leads to a surface impedance of the form

\bigskip

\[
\hfill\hfill\hfill \hfill\hfill \hfill  \hfill  Z_S  = \sqrt
{\frac{{\mu _o }}{{4\pi \varepsilon }}\frac{{\omega ^2 }}{{\left(
{\omega ^2 - \omega _p^2 } \right) + i\Gamma \omega }}} \hfill,
\hfill \hfill \hfill \hfill \hfill (2) \hfill
\]

\bigskip

\noindent{where $\Gamma$ $(=\sigma_q/\epsilon)$ is a damping term
that depends only on the normal quasiparticle contribution to the
conductivity, $\sigma_q$. Eq. (2) gives rise to an asymmetric JPR,
as shown in Fig. 1. This particular simulation assumes
$\omega_p^2$ $\propto$ {\bf B}$^{-1}$, as determined from
experiment (see section III). Furthermore, in order to reproduce
the increased damping of the JPR with increasing field, as
observed experimentally, we assume a quasiparticle conductivity
which increases linearly with the flux density {\bf B}. The data
in Fig. 1 reproduce the main features observed from our previously
published measurements of the JPR frequency dependence
\cite{JPR1}, {\em i.e.} the asymmetric resonance broadens and
moves to higher field as the measurement frequency is reduced. The
reason for the shift in the resonance position is discussed
below.}

While our simulations attribute the broadening and asymmetry of
the JPR entirely to quasiparticle damping effects, there does
exist another possibility that has recently been discussed in the
literature \cite{Koshelev99}. Inhomogeneous broadening of the JPR
line, caused by random fluctuations in the interlayer Josephson
coupling, has also been shown to produce asymmetry in the JPR
line. We save a detailed discussion of this effect until the
latter sections of this article. The main purpose of the
simulation in Fig. 1 is to emphasize that the JPR asymmetry arises
naturally from electrodynamics, {\em i.e.} the measured
dissipation within the cavity is governed both by the real and
imaginary parts of the complex conductivity \cite{lambda}. Even if
one assumes a field independent dissipative mechanism, the
electrodynamics still result in an asymmetric JPR lineshape
similar to the one shown in Fig. 1.

%while the increased damping with increasing field is due to the
%greater number of vortices in the sample, which enhances the
%normal quasiparticle conductivity.

All microwave measurements were conducted in a mode where the
measurement frequency is held constant (due to the narrow band
technique), and the magnetic field is swept at different fixed
temperatures, {\em i.e.}, the field tunes $\omega_p$, and a JPR is
observed whenever $\omega_p(${\bf B},T) matches the measurement
frequency $\omega$. Application of a magnetic field generally
suppresses the critical current density along the {\em a}-axis,
thereby reducing $\omega_p$. This is due to the fact that an
increasing flux density, when combined with disorder and thermal
fluctuations in the vortex positions, tends to suppress interlayer
Josephson coupling. In a fixed frequency/swept field experiment,
therefore, any external factors that increase $\omega_p$ (e.g. a change in
temperature) will shift the observed JPR to higher magnetic field,
since a stronger field will be required to shift $\omega_p(${\bf
B}) down to the measurement frequency $\omega$. Conversely, any
external factors which reduce $\omega_p$ will shift the JPR to lower field.
This will be important for subsequent analysis of the temperature
dependence of the JPR in the following section, and accounts for
the shift in the resonance positions in Fig. 1, {\em i.e.}
stronger fields are necessary to suppress $\omega_p$ to lower
measurement frequencies.

Several different single crystals of
$\kappa-$(BEDT-TTF)$_2$Cu(NCS)$_2$, with approximate dimensions
$0.75\times0.5\times0.2$ mm$^3$, were used in this study; all of
the samples were grown in the same batch using standard techniques
\cite{Ishiguro}. We found that all samples gave qualitatively
similar results. Temperature control was achieved using a Cernox
thermometer and a small resistive heater attached mechanically to
the cavity \cite{mmrsi}. DC magnetic fields were applied parallel
to the sample's {\em a}-axis for all measurements, and field
sweeps were made at a constant rate of approximately $\pm 1$ T per
minute.

\bigskip

\centerline{\bf III. Results and discussion}

\bigskip

\noindent{Figure 2 shows temperature dependent microwave
dissipation ($\propto R_S$) for two different frequencies: a) 111
GHz and b) 134 GHz. The JPR is observed as a broad asymmetric
resonance (see Fig. 1 for comparison) with a strongly temperature
dependent amplitude and width. For these two frequencies, the
resonance peak position exhibits a non-monotonic dependence on
temperature. The top traces in each figure were obtained at the
lowest temperature ($\sim 2$ K), and the bottom traces at the
highest temperature ($\sim 10$ K) $-$ see figure caption for the
exact temperatures. As noted above, $J_m(${\bf B},T) and,
therefore $\omega_p$, is suppressed upon application of a magnetic
field, {\em i.e.} the JPR at the lower frequency of 111 GHz is
observed at higher fields. The higher field 111 GHz resonances are
also broader, as noted in the previous section. Figure 3 plots the
resonance positions, as a function of temperature, for
measurements at several different frequencies. The data clearly
define a line (dashed curve) separating two regime, one for which
$\partial\omega_p/\partial$T$<0$, the other for which
$\partial\omega_p/\partial$T$>0$. Similar behavior has been
observed in Bi$_2$Sr$_2$CaCu$_2$O$_{8+\delta}$
\cite{Tsui,Matsuda}. We note from above that a resonance which
moves to lower fields implies a reduction in $\omega_p$, and vice
versa.

Observation of a region of field/temperature parameter space for
which $\partial\omega_p/\partial$T$>0$, implies that Josephson
coupling is enhanced upon raising the temperature. Or, phrased in
another way, this implies that raising the temperature results in
an increased correlation in the positions of pancake vortices in
adjacent layers. This rather counter-intuitive result can only be
understood in terms of a vortex state which exhibits some degree
of pinning \cite{Bul96,Matsuda97}. It is already well established
that the 3D flux-line-lattice decouples at relatively weak fields
on the order 10 mT \cite{Lee}. However, the nature of the
subsequent vortex state has not been well established, since the
loss of a 3D ordered state renders most techniques ({\em e.g.}
$\mu$SR \cite{Lee}) insensitive to any remaining long range order
within the layers. One possible reason for the enhancement in the
quantity $\langle\langle\cos\varphi_{n,n+1}(${\bf
r}$)\rangle_t\rangle_d$ ($\propto\omega_p$), upon increasing the
temperature, is illustrated by means of the schematic in Fig. 4.
Vortex-vortex (1/{\bf r}) interactions result in intra-layer
correlations in the locations of vortices within each particular
layer. A Q2D hexagonal vortex lattice (as depicted in Fig.4)
represents a limiting case of this correlated state, though a Q2D
glassy state probably offers a more realistic description of the
apparent pinned vortex phase discussed here. Defects (not shown)
pin a small fraction of the vortices. Since the intra-layer
vortex-vortex interactions overwhelm the interlayer Josephson
coupling at fields above the decoupling field (few mT), collective
pinning has the effect of locking the positions of vortices in one
layer independently of the positions of the vortices in adjacent
layers. Consequently, pancake vortices in layers {\em n} and
$n-1$, which are threaded by the same flux quantum, do not
necessarily occupy the same position in the $xy-$plane. This leads
to a finite separation ($r_{n,n-1}$) of the {\em xy}-coordinates
of vortices in layers {\em n} and $n-1$, and contributes to a
suppression of $\langle\langle\cos\varphi_{n,n+1}(${\bf
r}$)\rangle_t\rangle_d$. Interlayer Josephson coupling provides a
weak restoring force which acts to restore linearity among the
pancake vortices. The combined potentials due to inter-vortex
repulsion (deep narrow minimum in Fig. 4b $-$ dashed curve) and
Josephson coupling (broad shallow minimum in Fig. 4b $-$ dashed
curve), results in an asymmetric (anharmonic) potential minimum
(solid curve in Fig. 4b) for the separation, $r_{n,n-1}$, between
vortices in layers {\em n} and $n-1$. One can now see that
increased thermal fluctuations will lead to a reduction in
$\langle\langle\varphi_{n,n+1}(${\bf r}$)\rangle_t\rangle_d$ and,
hence, an increase in $\langle\langle\cos\varphi_{n,n+1}(${\bf
r}$)\rangle_t\rangle_d$, {\em i.e.} thermal fluctuations lead to
increased linearity of the flux lines.

One other possible explanation for the
$\partial\omega_p/\partial$T $>0$ behavior, which has been
discussed by Matsuda {\em et al}. in connection with JPR
measurements on
Bi$_2$Sr$_2$CaCu$_2$O$_{8+\delta}$\cite{Matsuda97}, involves a
non-equilibrium critical state. We note that the existence of a
critical state implies collective pinning and glassy behavior,
{\em i.e.} pinning plus intra-layer vortex correlations. In this
picture, in-plane currents associated with the critical state
exert Lorentz forces on the pancake vortices. For situations in
which the interlayer Josephson coupling is weak, as is the case in
our experiments, the Lorentz forces drive the pancake vortices out
of alignment in a somewhat analogous fashion to the collective
pinning scenario discussed above. Consequently, interlayer
coherence is suppressed. Furthermore, this suppression of
$\langle\langle\cos\varphi_{n,n+1}(${\bf r}$)\rangle_t\rangle_d$
increases upon lowering the temperature, since the in-plane
critical currents increase with decreasing temperature. Field
cooled experiments offer a means of distinguishing between these
two scenarios. However, for the purpose of this article, we note
that both explanations result in the same conclusions concerning
the nature of the vortex state in the regions of Fig. 3 where
$\partial\omega_p/\partial$T $>0$.

Having established the existence of a Q2D pinned vortex phase over
a considerable portion of the mixed state phase diagram for the
title compound, one next has to consider the reasons for a
crossover in the temperature dependence of the JPR frequency
(dashed line in Fig. 3). $\partial\omega_p/\partial$T$<0$ implies
a decreasing interlayer coherence with increasing temperature. The
first possible explanation is a melting or glass transition
\cite{Zeldov,Schilling,Houghton,Fruchter}, whereby thermally or
quantum induced fluctuations in the positions of the pancake
vortices become comparable to the average in-plane inter-vortex
separation \cite{Lindemann}. In this scenario, pinned vortices may
remain pinned; however, long range order among the vortices is
completely lost. The result is a Q2D flux liquid state in which
inter-vortex repulsion plays a much reduced role. Nevertheless,
the liquid state may exhibit some residual viscosity due to the
residual inter-vortex interactions between mobile and pinned
vortices \cite{BlatterRMP}. The transition may be expressed in
terms of either a temperature dependent critical field, {\bf
B}$_m$(T), or as a field dependent critical temperature,
T$_m$({\bf B}). Once the fluctuations in the positions of pancake
vortices become large in comparison to the average in-plane vortex
separation [{\em i.e}. T$>$T$_m$({\bf B}) or {\bf B}$>${\bf
B}$_m$(T)], the model described above to explain
$\partial\omega_p/\partial$T$>0$ breaks down (see Fig. 4). In this
limit, increased thermal fluctuations serve only to suppress
$\langle\langle\cos\varphi_{n,n+1}(${\bf r}$)\rangle_t\rangle_d$;
hence, the observed temperature dependence of $\omega_p$.

A subtly different explanation for the observed crossover in the
temperature dependence of the JPR frequency (dashed line in fig.
3) involves a depinning transition \cite{depin1,depin2}. In this
picture, pinning forces are considerably weaker than the
inter-vortex repulsive forces that stabilize the correlated state.
Upon raising the temperature, vortices execute larger and larger
collective displacements from equilibrium. In fact, this
collective behavior would be indistinguishable from the picture
described above to explain $\partial\omega_p/\partial$T$>0$ (see
Fig. 4), {\em i.e.} the collective fluctuations would lead to an
increase in $\langle\langle\cos\varphi_{n,n+1}(${\bf
r}$)\rangle_t\rangle_d$, as observed for T$<$T$_m$({\bf B}). Upon
exceeding a critical depinning threshold (characteristic of the
nature of the pinning), the vortex arrays become completely
depinned or mobile. The result is a depinned or mobile Q2D vortex
state in which pinning forces play a much reduced role. As in the
liquid case, the depinned phase may be expected to exhibit a
finite viscosity due to the weak residual interaction with pinning
centers. Furthermore, once the collective displacements of the
vortices exceed the average in-plane vortex separation, one
expects the temperature dependence of $\omega_p$ to crossover in
the same way as described above for the glass/liquid transition.

To reiterate, the glass/liquid and depinning transitions seem to
describe qualitatively similar phenomena. However, the physical
origin of each transition is rather different. In the former case,
thermal fluctuations overcome the inter-vortex interactions that
maintain correlations among vortices within each layer, while in
the depinning case, thermal fluctuations overcome the pinning
forces and the correlated flux bundles become mobile.
Unfortunately, because the JPR frequency depends only on the
averaged value of the $\cos \varphi_{n,n-1}(${\bf r}$)$, it is
only sensitive to global changes in the critical current density,
and not the mechanisms for such changes. However, for either case
mentioned above, the opposing temperature dependence observed
above and below the transition temperature T$_m$({\bf B}), should
be expected.

Having identified a transformation in the vortex structure from
the temperature dependence of the JPR frequency, we next turn our
attention to the field dependence of $\omega_p$ in each phase. Our
previous efforts to fit $\omega_p$ versus {\em B} were complicated
by the fact that insufficient data were obtained completely within
each of the phases \cite{JPR1}, {\em i.e.} much of the data
straddled the transition line. Various theoretical models predict
quite different behavior for ordered and disordered phases, as
well as a strong dependence on the nature of the pinning.
Recently, A. E. Koshelev has derived the explicit form of
$\omega_p$ within the Q2D vortex liquid state, using field
theoretical methods \cite{Koshelev98}:

\[
\hfill\hfill\hfill\hfill\hfill\hfill\hfill\omega _p^2  =
\frac{{4\pi dJ_o^2 \Phi _o }}{{\varepsilon k_B {\rm TB}}}
\hfill;\hfill\hfill\hfill\hfill \hfill (3) \hfill
\]

\noindent{here, {\em d} is the interlayer separation, and $\Phi_o$
is the flux quantum. Consequently, within the Q2D vortex liquid
state, the JPR frequency should follow a $\omega_p^2 \propto
1/$T{\bf B} dependence. To this end, $\omega_p^2$ is plotted as a
function of applied field in Fig. 5. These data points were all
obtained at 7 K and at fields above the transition field {\bf
B}$_m$(T), {\em i.e.} in the disordered state. The solid line is a
powerlaw fit of the form $\omega_p^2 =$ A{\bf B}$^{-\mu}$, with
$\mu = 0.93 \pm 0.05$. Since this value is close to unity, it
supports the assumption that the system is in a highly disordered,
or liquid-like state. The value of $J_o = 2 \times 10^7$ Am$^{-2}$
obtained from the fit is in excellent agreement with the value
obtained from previous investigations \cite{JPR1}.

Within the low-T/low-{\em B} phase, determination of the field and
temperature dependence of the resonance frequency gives insight
into the nature of the pinning \cite{Bul95,Bul96}. Like the vortex
liquid state, the JPR frequency is expected to follow a powerlaw
dependence, with an exponent that tends to unity with increasing
disorder. A reduction in the number of pinning centers in cleaner
crystals results in a better alignment of the pancake vortices.
This, in turn, leads to a reduction in the magnitude of the
exponent in the powerlaw, {\em i.e.} to a much slower suppression
of $\omega_p$ with field. Fig. 6 plots the field dependence of the
JPR frequency for data obtained entirely within pinned vortex
glass phase, {\em i.e.} for fields below the transition field {\bf
B}$_m$(T). The solid curve is a powerlaw fit to the data, from
which an exponent of $\mu = 0.35\pm 0.02$ is obtained, indicative
of an intermediate to low degree of pinning within this phase. We
note that similar experiments on several HTS compounds also show a
powerlaw behavior, with exponents of order unity in every case
\cite{Tsui,Matsuda}. This is seen even within the ordered vortex
solid state, indicating that the materials studied here are
considerably cleaner than typical HTS materials, something which
has been well established via a range of other techniques.
Batch-to-batch variations in sample quality have, however, been
noted for this compound. For example, there is no overlap between
the data in Fig. 6 and data from Fig. 5 of ref. \cite{JPR1}, which
were obtained for a sample from a different synthesis.
Furthermore, several other groups have published JPR data for the
title compound, and there is rarely good agreement between the JPR
peak positions for a given frequency and temperature
\cite{Shib1,Shib2,Comment,Schrama}.

%In the limit of extremely weak pinning, Bulaevskii {\em et al.}
%have shown that when the average deviation of vortices away from
%linearity is less than the superconducting coherence length, the
%JPR frequency should exhibit a stretched exponential dependence on
%the magnetic field of the form,

%\[ \hfill \hfill \hfill \hfill \hfill \hfill \hfill
%\omega _p^2  = \frac{{\pi d}}{{\varepsilon _o \Phi _o }}J_o \exp
%\left[ { - \left( {{\rm B/B}_D } \right)^{{3 \mathord{\left/
% {\vphantom {3 2}} \right.
% \kern-\nulldelimiterspace} 2}} } \right] \hfill, \hfill \hfill \hfill \hfill
% \hfill (4) \hfill
%\]

%\noindent{where {\bf B}$_D$
%($=\Phi_o^{11/3}/[(4\pi)^3(2\pi\gamma_o {\rm
%E}_p)^{2/3}d^2\lambda_{\|}^4]$) is a decoupling field, above which
%interlayer coherence is lost, E$_p$ is a pinning parameter which
%measures the amount of disorder in the sample, and $\gamma_o$ is
%the ratio of the interlayer and in-plane penetration depths
%($\lambda_\bot/\lambda_\|$) in zero field \cite{Bul95,Bul96}.}

%A similar plot is shown in the main part of the figure where,
%instead, $\omega_p^2$ is graphed against {\bf B}$^{3/2}$; the
%solid line is a fit to Eq. (4), giving a value of {\bf B}$_D =
%0.15$ T. Both Eq. (4) and the powerlaw fit produce reasonable
%agreement with the data, with comparable $\chi^2$ values.
%Therefore, it seems that the measured crystal sits right at the
%borderline between a moderate to extremely weakly pinned Q2D VL.

Finally, we comment on the linewidth and shape of the JPR. While
the model discussed in ref. \cite{Koshelev99} (involving
inhomogeneous broadening due to random Josephson coupling $-$ see
also \cite{Bul96r2}), accounts for the observed asymmetry of the
JPR, it cannot account for the field dependence of the JPR widths
observed from our measurements. In particular, this model predicts
a linewidth which is inversely proportional to magnetic field
strength. Examination of Figs 2a) and b) reveal a completely
opposite trend. Indeed, the JPR width is approximately
proportional to {\bf B}, {\em i.e.} the data in Fig. 6a), which
are observed at roughly twice the field strength of the data in
Fig. 6b), span approximately twice the field window of the data in
Fig. 6b). This trend is even more apparent from our earlier
studies (see {\em e.g.} Fig. 2 of ref. \cite{JPR1}). Although we
cannot completely rule out an inhomogeneous contribution to the
linewidths, the pronounced broadening of the JPR with field
suggests that quasiparticle damping effects dominate both the line
widths and shapes.

\bigskip

\centerline{\bf IV. Summary and conclusions}

\bigskip

\noindent{We have utilized a resonant cavity perturbation
technique to probe the interlayer electrodynamics of the Q2D
$\kappa-$(BEDT$-$TTF)$_2$Cu(NCS)$_2$ organic superconductor. A JPR
is observed, which proves to be extremely sensitive to
correlations in the locations of vortices in adjacent layers. By
following the temperature and field dependence of JPR frequency
($\omega_p$) at many frequencies, a clear transition line emerges
near the irreversibility line, where a first order phase
transition has also been observed from magnetic measurements
\cite{Inada}. A global mixed state phase diagram, which includes
this data, has recently been published in refs.
\cite{MolaPRL,Single2}. Below the transition line {\bf B}$<${\bf
B}$_m$(T) [or T$<$T$_m$({\bf B})], the existence of a correlated
Q2D vortex (glassy) phase has been established. Two possibilities
have been considered for the vortex transformation that takes
place at {\bf B}$_m$(T) $-$ namely, melting and depinning
transitions. In either scenario, the high-{\bf B}/high-T state
resembles a liquid-like state (up to {\bf H}$_{c2}$). In the
melting scenario, residual inter-vortex interactions cause some
remnant local order or clustering of vortices around pinning
sites, but longer range correlations are suppressed
\cite{BlatterRMP}. Under the depinning scenario, finite vortex
correlations exists; however, the 2D vortex arrays may break up
into domains or flux bundles, with some still pinned, while others
are mobile. This represents a so-called vortex slush phase.
Similar plastic flows of vortices have been observed in the HTS
\cite{Reyes}. Essentially, the vortex slush and the pinned liquid
are the same, with the subtle difference being the inter-vortex
correlation length, {\em i.e.} the sizes of the pinned and mobile
domains is much larger for the slush than for the pinned liquid.
It is clear from these investigations, that organic
superconductors offer the opportunity to investigate a distinctly
different pinning regime from the more widely studied HTS
materials.}

\bigskip

\centerline{\bf V. Acknowledgements}

\bigskip

\noindent{This work was supported by the National Science
Foundation (DMR0196461 and DMR0196430). S. H. is a Cottrell
Scholar of the Research Corporation.}

%\end{multicols}

\clearpage

\noindent{\bf Figure captions}

\bigskip

Fig. 1. Simulations of the frequency and magnetic field dependence
of the surface resistance given by Eq. (2). The asymmetric peak is
due to the JPR.

\bigskip

Fig. 2. Temperature dependence of the field dependent microwave
dissipation observed at a) 111 GHz and b) 134 GHz. The asymmetric
peak is due to the JPR (see Fig. 1 for comparison). The
temperatures in each figure are, from top to bottom (in kelvin):
2.0, 2.2, 2.5, 2.7, 3.1, 3.2, 3.5, 3.7, 4.0, 4.2, 4.5, 4.7, 5.0,
5.2, 5.5, 5.7, 6.0, 7.0, 8.0, 9.0.

\bigskip

Fig. 3. A compilation of the temperature dependence of the peak
positions of the JPR observed at several different frequencies
(indicated in the figure). The dashed line separates regions with
different temperature dependencies for $\omega_p$.

\bigskip

Fig. 4 a) Schematic representing a Q2D vortex solid, in which long
range hexagonal order exists within each layer, yet there is no
correlation in the locations of vortices in adjacent layers, even
though they may be threaded by the same flux quantum. b) The
resultant potential U($r_{n,n-1}$) (solid line) for a vortex in
layer $n-1$ subjected to a deep potential minimum (dashed curve)
due to the rigidity of the vortex lattice in that layer, and a
weaker offset potential minimum (due to the interlayer Josephson
coupling) which acts to line up pancake vortices threaded by the
same flux quantum.

\bigskip

Fig. 5. Magnetic field dependence of the squared JPR frequency
within the depinned or liquid state (T = 7 K). The solid line is a
powerlaw fit to the data.

\bigskip

Fig. 6. Magnetic field dependence of the squared JPR frequency
within the weakly-pinned Q2D ordered/glassy phase (T = 2 K). The
solid curve shows the fit to a powerlaw [Eq. (3)].

\clearpage

\begin{figure}
\centerline{\epsfig{figure=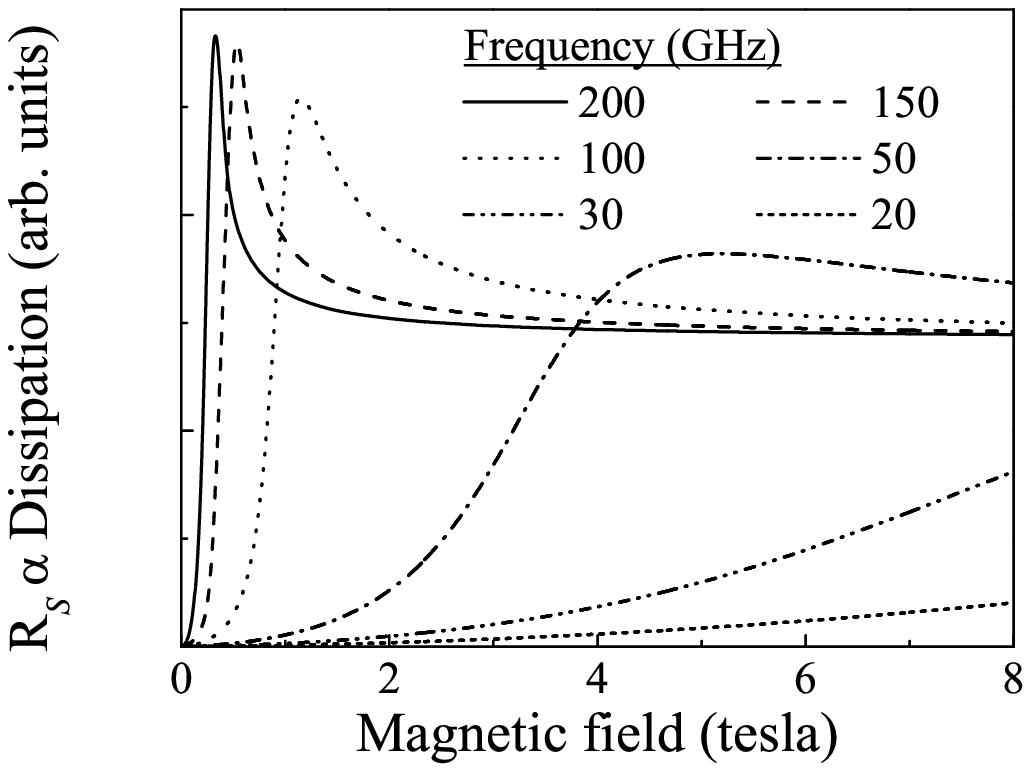,width=120mm}}
\bigskip
\caption{Hill {\em et al.}} \label{Fig. 1}
\end{figure}

\clearpage

\begin{figure}
\centerline{\epsfig{figure=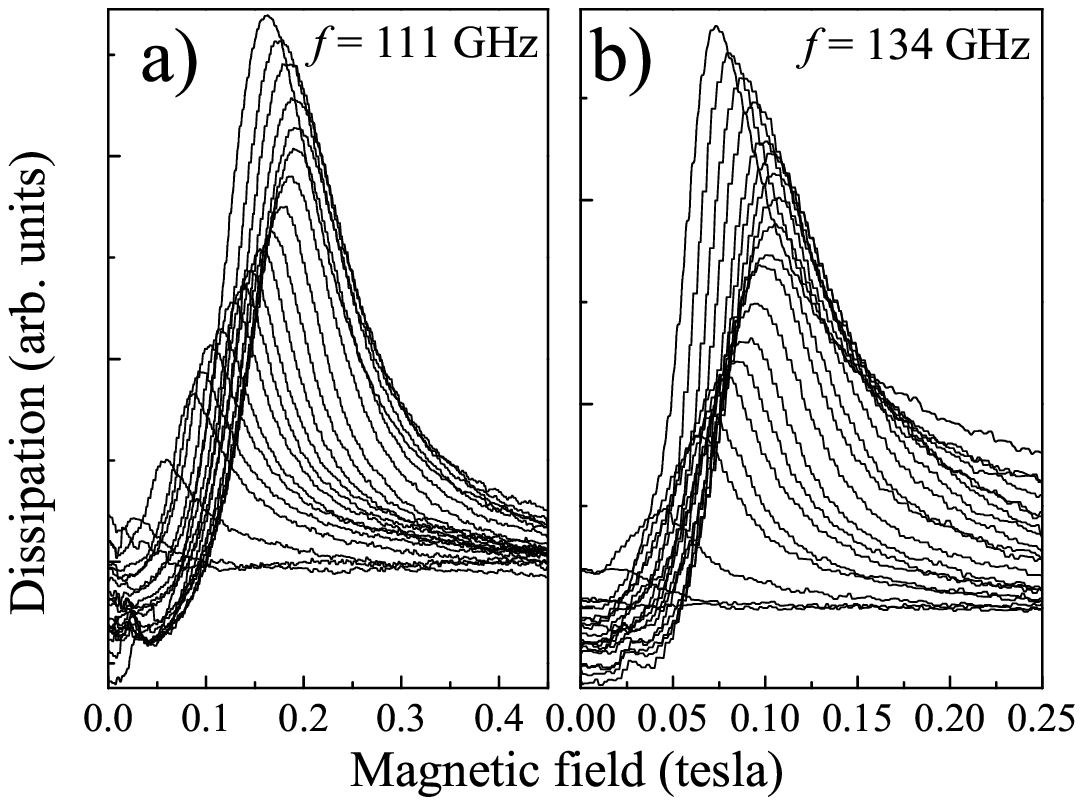,width=140mm}}
\bigskip
\caption{Hill {\em et al.}} \label{Fig. 2}
\end{figure}

\clearpage

\begin{figure}
\centerline{\epsfig{figure=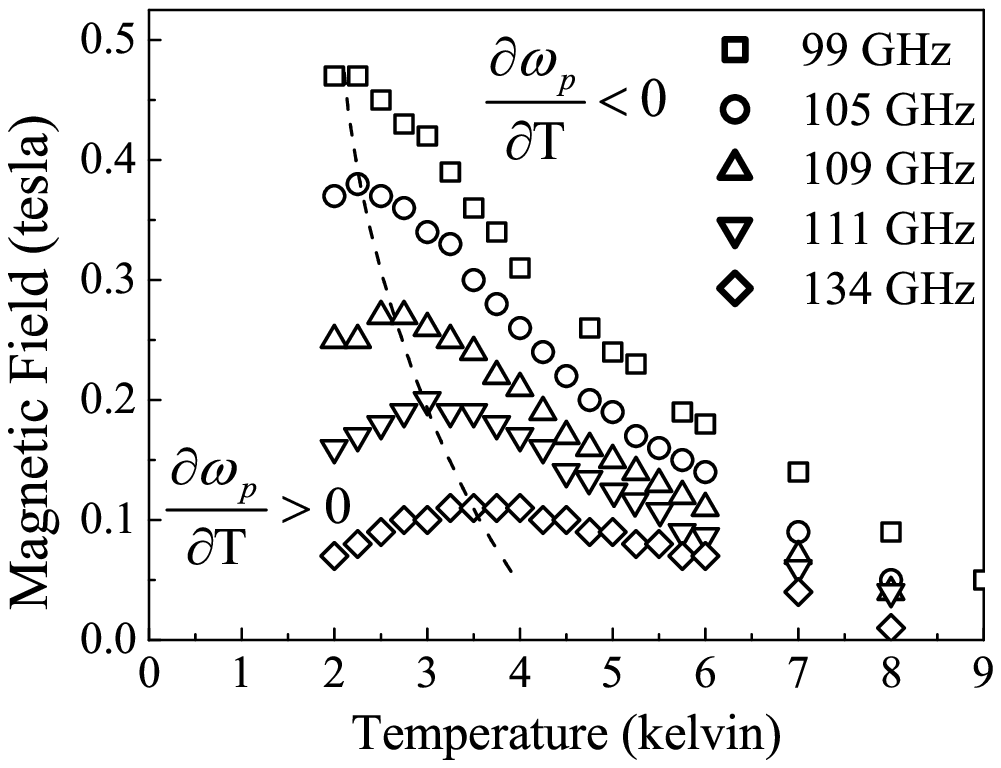,width=120mm}}
\bigskip
\caption{Hill {\em et al.}} \label{Fig. 3}
\end{figure}

\clearpage

\begin{figure}
\centerline{\epsfig{figure=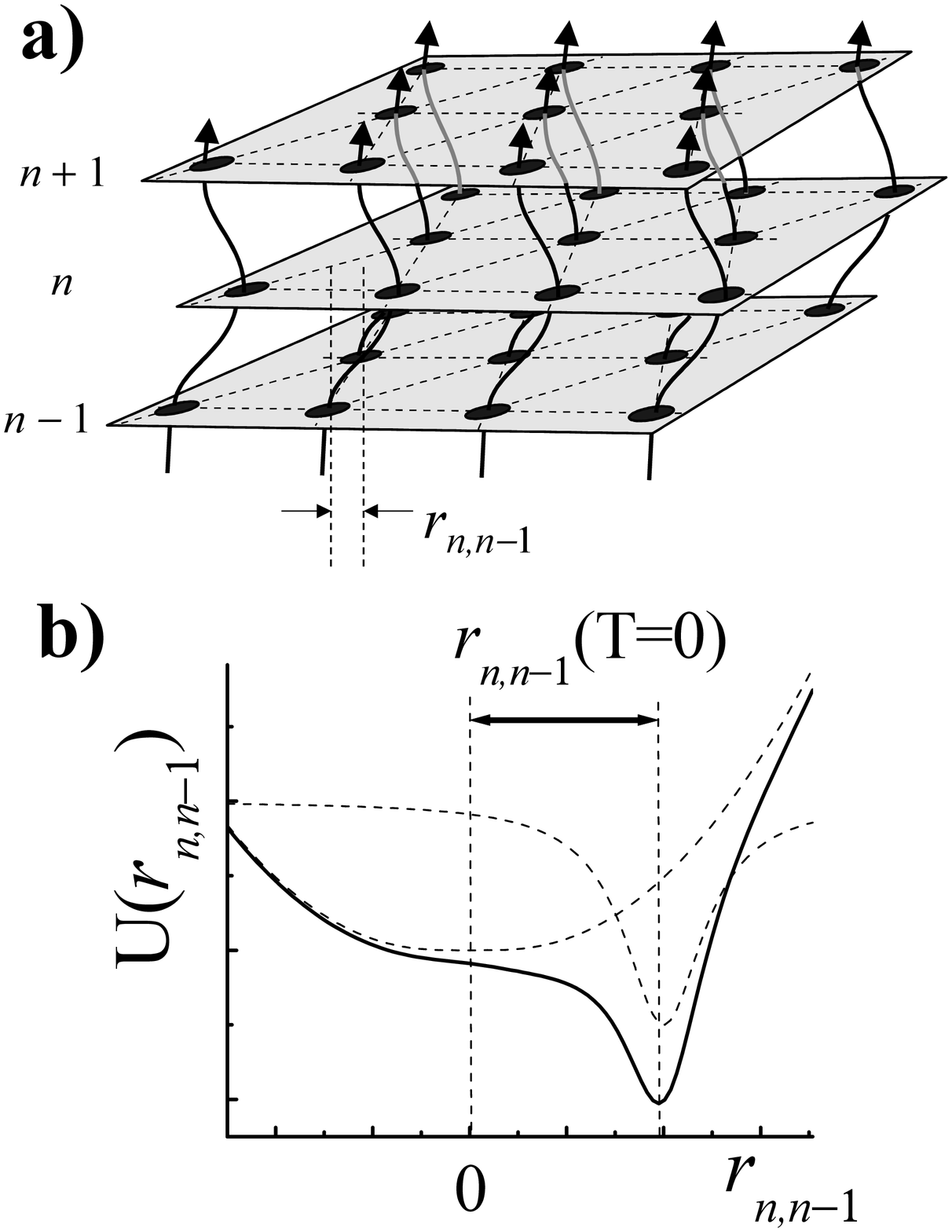,width=90mm}}
\bigskip
\caption{Hill {\em et al.}} \label{Fig. 4}
\end{figure}

\clearpage

\begin{figure}
\centerline{\epsfig{figure=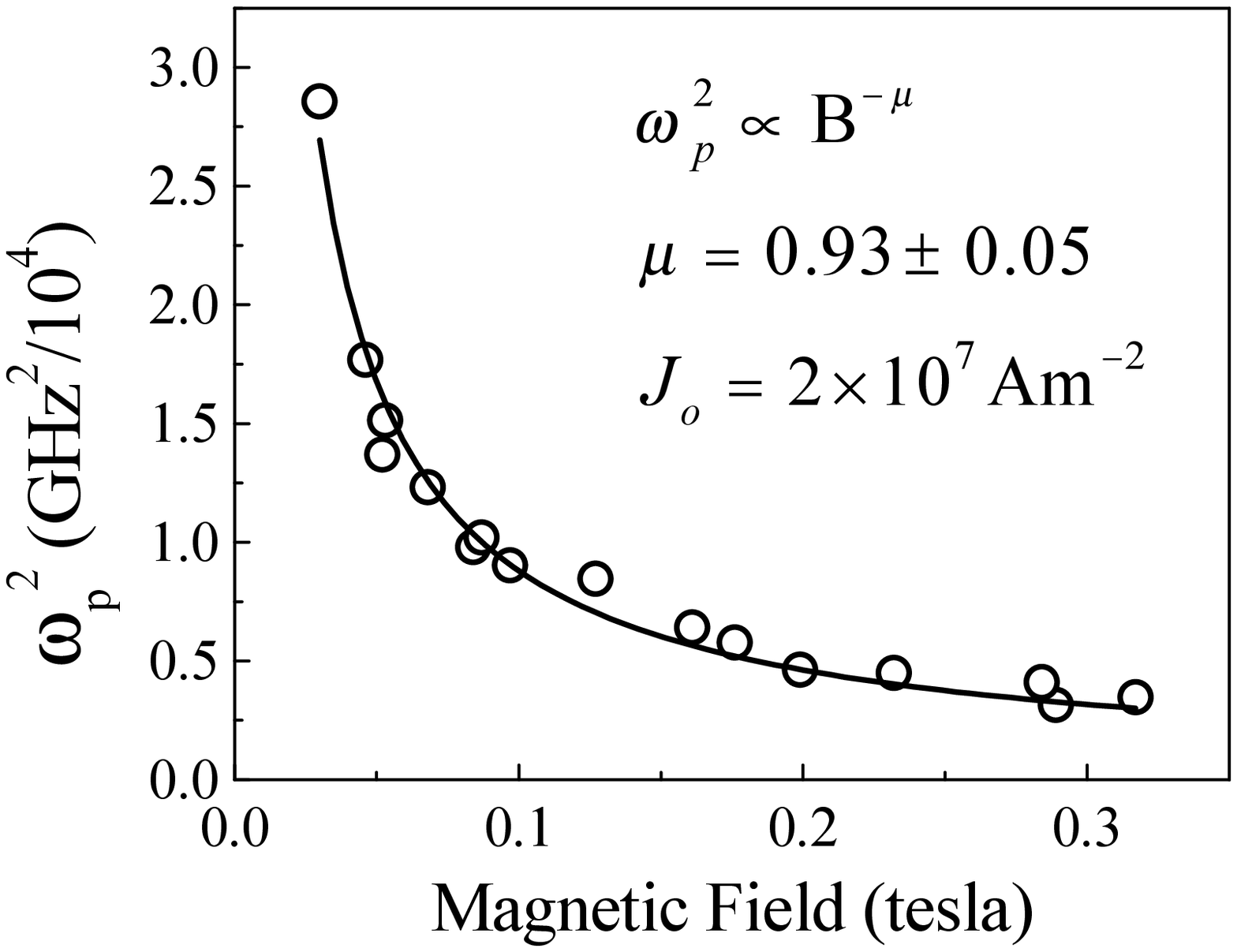,width=140mm}}
\bigskip
\caption{Hill {\em et al.}} \label{Fig. 5}
\end{figure}

\clearpage

\begin{figure}
\centerline{\epsfig{figure=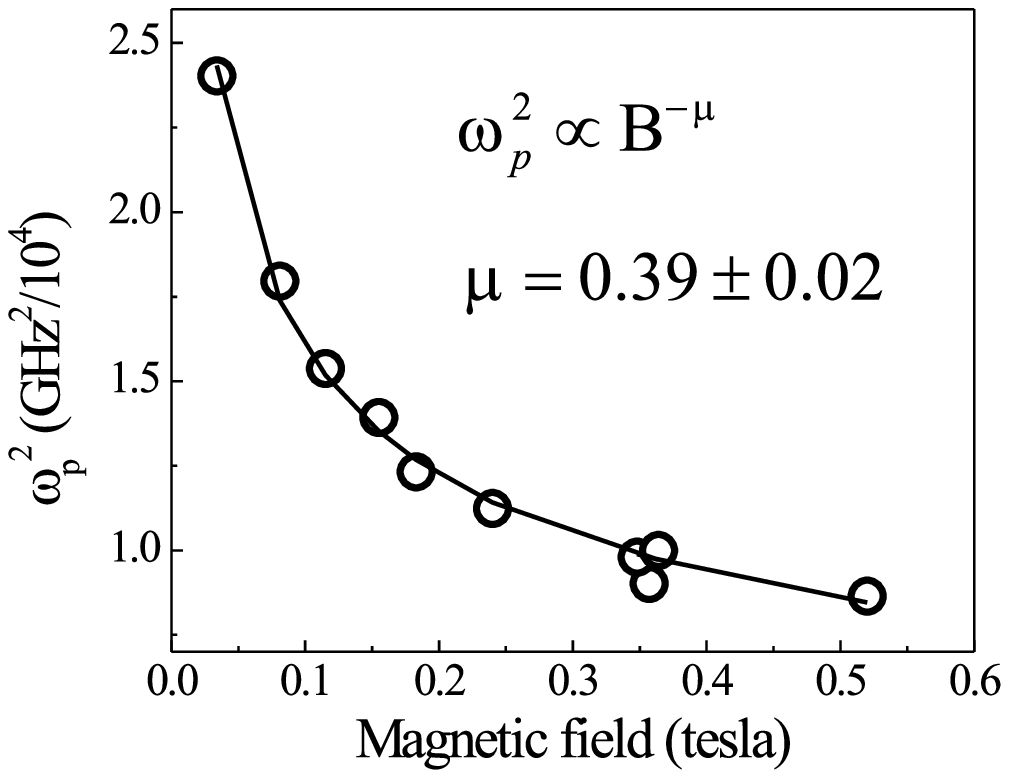,width=140mm}}
\bigskip
\caption{Hill {\em et al.}} \label{Fig. 6}
\end{figure}

\clearpage

\end{document}